\documentclass[12pt]{article}
\usepackage{amssymb}
\usepackage{amsfonts}
\def\dspace{\baselineskip = .30in}

\def\beq{\begin{equation}}
\def\eeq{\end{equation}}

\begin{document}
\begin{titlepage}

\begin{center}
{\Large\bf  D-brane Inflation\footnote{Talk given by Q.S. at the workshop
in
Heidelberg (April 4-7, 2001),
and at the EURESCO conference in Le Londe (France) May 11-16, 2001.}}
\end{center}
\vspace{0.5cm}
\begin{center}

{\large G. Dvali$^{a}$\footnote {E-mail address:
gd23@scires.nyu.edu} {}~and
{}~Q. Shafi$^{b}$\footnote {E-mail address:
shafi@bartol.udel.edu} and S. Solganik$^{a}$\footnote{E-mail address:
ss706@scires.nyu.edu}}
\vspace{0.5cm}

$^a${\em New York University, Department of Physics, New York, NY  10003,
USA.\\

$^b$ Bartol Research Institute, University of Delaware, Newark, DE  19716,
USA.}\\
\end{center}

\hspace{1.0cm}

\begin{abstract}

We discuss a calculable version of brane inflation, in which a set of
parallel D-brane and anti-D-brane worlds, initially displaced in extra
dimension, slowly attract each other. In the effective four-dimensional
theory this slow motion of branes translates into a slow-roll of a scalar
field
(proportional to their separation) with a flat potential that drives
inflation. The number of possible e-foldings is severely constrained. The
scalar spectral index is found to be 0.97, while the
effective compactification scale is of  order  $10^{12}$ GeV. Reheating of
the Universe is provided by collision and subsequent annihilation of
branes.

\end{abstract}
\end{titlepage}
\newpage
\dspace
\bigskip

An inflationary scenario may be regarded as successful if it
satisfies the following constraints:

$(i)$ The vacuum energy V that drives inflation is sufficiently
flat along the inflationary trajectory, so that at least the
required minimum number (say 60) of e-foldings can be generated
to resolve the horizon and flatness problems.

$(ii)$ The spectral index of the scalar density fluctuations
should be close to unity (see, for instance, \cite{cmb}).
Of course, the magnitude of
${\delta}T/T$
generated by inflation should also agree with the
observations.

$(iii)$ A mechanism should exist for providing a satisfactory
end to inflation and subsequent generation of the observed
baryon asymmetry.

Among all these constraints, it appears that $(i)$ is often the
hardest to satisfy.
The `slow roll' condition, needed
to implement inflation, requires that the effective interaction
term
$(V(\phi)/M^{2}_{P})\phi^{*}\phi$
be absent along the inflationary trajectory.  Here $\phi$ denotes
the inflaton (scalar field that drives inflation), and
$M_{P}=2.4\times10^{18}$ GeV is the reduced Planck mass.
The absence of this term is demanded so that
$m^{2}_{\phi}{\ll}H^{2}$,
and inflation becomes possible.
This is
hard to understand in conventional four-dimensional theories.
In general, it is expected that quantum gravity corrections can spoil
this flatness. If the Hubble parameter during the
`would' be inflation is $H$, it is generally believed that the quantum
gravity correction can generate
a curvature $\sim H^2$
of the inflaton potential, and thus violate the slow-roll conditions
necessary for the inflation.

To make the origin of this correction
clearer,
let us assume that the potential that can lead
to the successful slow-roll conditions is
$V(\phi)$, where $\phi$ is an inflaton field. One of the necessary
conditions
is that the curvature of the potential, at least in some region, is smaller
than
the Hubble parameter
\begin{equation}
 H^2 \sim V/3M_P^2.
\label{sloroll}
\end{equation}
Suppose this is the case for a given $V$. However, it is hard to
understand
what forbids in the low-energy effective theory terms such as
\begin{equation}
\bar{\phi}\phi{V \over M_P^2}.
\label{corrections}
\end{equation}
(bar stands for hermitian conjugation).
These are not forbidden by any symmetries of the effective field theory.
If present, they would break condition
(\ref{sloroll}), unless various parts of
the potential are carefully adjusted.

A nice example of the difficulty involved is provided by the
following relatively well motivated inflationary model based on
supersymmetry
(SUSY) \cite{shafi}, which in fact is a supersymmetric realization of the
hybrid inflationary scenario\cite{linde}.  Consider the symmetry breaking
$G {\rightarrow}H$,
implemented through the superpotential
\beq
W={\kappa}S(\bar{\phi}\phi - \mu^{2}),
\eeq
where ${\kappa},{\mu}^{2}$
can be taken positive, $S$ is a
gauge singlet superfield, and
$\phi$ and its adjoint ${\bar\phi}$ belong to suitable
representations of $G$ in order to break it to $H$.
>From $W$ it is readily checked that the supersymmetric (SUSY)
minimum corresponds to the following vacuum expectation value:
$$|\langle {\phi}\rangle |=|\langle {\bar\phi}\rangle | =
{\mu},\;\;\;\langle S\rangle = 0.$$
(After SUSY breaking of order $m_{3/2}$, $\langle S\rangle$ may acquire a
value
proportional to $m_{3/2}$).

In order to implement inflation we must assume that in the early
universe, the fields are displayed from their present day minimum.
For  $S \gg{S_{c}}={\mu}$,
the minimum of the potential corresponds to
$\langle \phi\rangle = \langle\bar\phi\rangle = 0$.
In other words, there is symmetry restoration, while SUSY is
broken by the non-zero vacuum energy density
$\kappa^{2}\mu^{4}$
(It may help to think of the real part of the complex scalar $S$
as temperature).  Taking account of mass splittings in the
$\phi,\bar{\phi}$
supermultiplets, the one loop radiatively corrected superpotential
is given by
\beq
V_{eff}(S)=\kappa^{2}\mu^{4}\left[1+{\frac{d\kappa^{2}}{16\pi^{2}}}
\log\left(\frac{\kappa^{2}|S|^{2}}{\Lambda^{2}}\right)\right],
\eeq
where $d$=dimension of $\phi$, $\bar\phi$, and $\Lambda$ is some cutoff.

The scalar spectral index turns out to be
\beq
n\simeq 1-\frac{1}{N_{Q}}=0.98,
\eeq
where $N_{Q}$ denotes the number of e-foldings experienced during
inflation by the current horizon
scale.  The symmetry breaking scale can be estimated from
the following approximate formula for
$(\delta{T}/{T})_{Q}$ \cite{lyth}:
\beq
\left(
\frac{\delta T}{T}
\right)_{Q}\simeq \frac{{8}{\pi}}{\sqrt{d}}\left(\frac{N_{Q}}{45}\right)
\left(\frac{{\mu}}{M_{P}}\right)^{2},
\eeq
which yields
$\mu \sim 10^{15}-10^{16}$ GeV,
reminiscent of the grand unification scale.
This looks promising but the important challenge now is to
address the supergravity (SUGRA) corrections.  Do the latter
generate a mass$^{2}$ term for $S$ along the inflationary
trajectory that ruins inflation?  Within the SUGRA
framework, the potential takes the form
\beq
V=\exp(K/(M^{2}_{P})\left[(K^{-1}){_{i}^{j}} F^{i}
F_{j}-\frac{{3}{|W|^{2}}}{{M^{2}_{P}}}
\right],
\eeq
where the Kahler potential $K$ has the form
\beq
K=|S|^{2} + |\phi|^{2}+|{\bar\phi}|^{2}+\frac{{b}|S|^{4}}{M^{2}_{P}}+ ...,
\eeq
and
$F^{i} = W^{i} + K^{i}W/M^{2}_{P}$,
with upper (lower)
indices denoting differentiation with respect to
$\phi_{i}$ (${\phi}^{j*}$).

The $|S|^{2}$ term in $K$ could prove troublesome, but
fortunately with the form of $W$ we have chosen (which can be
generalized to include higher order terms, using a suitable
$U(1)$ R-symmetry), the problem with the mass $^{2}$ term for
$S$ is evaded.  However, the $|S|^{4}$ term in $K$ generates,
via
${\partial^{2}K}/\partial S\partial S^{\*}=1+4b|S|^{2}/M^{2}_{P}$,
a mass$^{2}$ term for S of order $H^{2}$ which
is bad for inflation!
Thus $b\ll 1$ (say of order
$10^{-3}$ or less)
is needed in order to implement inflation,
which may not seem very appealing.  Fortunately, the higher order terms
in $K$ turn out to be harmless.

The above discussion provides a good example of how SUGRA corrections
can potentially ruin an otherwise quite respectable inflationary
scenario.  So the question to ask is:  How can we do better?

One way to avoid such terms is to rely on a more fundamental
theory that could forbid such terms due to reasons that cannot be seen
in the low energy effective field theory as the result of symmetries.
An example of such a situation is provided by the idea of
``brane inflation''\cite{dva}, in which the dangerous mass term for the
inflaton is forbidden
by locality of the high-dimensional theory.
In this picture, the inflaton is a field ($\phi$) that parameterizes
the distance
($r$)
between two brane worlds embedded in the extra space. The typical distance
between these two branes is much bigger than the string scale, so that the
potential between them is essentially governed by the infrared
bulk (super)gravity.  The effects of
higher string excitations are all decoupled.
The potential at
large distances has an inverse power-law
dependence on the inter-brane distance
\begin{equation}
V(r) = M^4_s(a   - {b \over (M_{s}r)^{N -2}})
\label{thepotential}
\end{equation}
where $M_s$ is the string scale, and $a$ and $b$ are some constants.
In the effective four-dimensional
field theory picture, this potential translates as the potential for the
inflaton field $\phi$ and is automatically sufficiently flat.
Thus, in this picture the inflation in four dimensions in nothing but
the brane motion in the extra space. Branes falling on top of each other
drive inflation in our space. The flatness of the inflaton potential is
the result of the locality in high-dimensional theory: no terms with
positive powers
of $\phi$ can be generated in the effective potential, due to the fact
that inter-brane interactions fall-off with the distance.

The aim of the present paper is to discuss a ``calculable'' version of the
brane inflationary scenario\cite{dva}, in which inflation is driven by
D-anti-D-brane pair
that attract each other and annihilate. The advantage of such a set-up is
that
the brane-antibrane potential is rather well known and has fewer free
parameters.
As we shall see this puts severe restriction on the number of e-foldings
during inflation. Before discussing this scenario let us briefly
review some needed properties of D-branes.
D-branes
are soliton like configurations that arise in type IIA, IIB and
type I string theories.  For instance, in type II B theory, the
elementary excitations are closed strings, while  D-p branes in
these theories are p-dimensional objects (p odd), whose dynamics
is described by the theory of open strings with ends
lying on the D-brane.  D-branes also carry Ramond-Ramond (RR)
charges, and their mass (tension) is proportional to 1/g, where g
denotes the string coupling.

The interaction energy between two parallel static D-p branes
in the large separation limit ($r \gg M_{s}^{-1}$, where $M_{s}$ denotes
the string scale)
is given  by \cite{bach}
\beq
{\cal
E}(r)=2V_{p}\kappa^{2}_{(10)}[{\rho^{2}_{(p)}}-{T^{2}_{(p)}}]\Delta^{E}_{(9-p)}(r),
\eeq
where
$\kappa_{(10)}=M_{10}^{-4}$ ($M_{10}$ is the ten-dimensional Planck mass),
$T_{(p)}$ is the brane tension,
$\rho_{(p)}$ is the RR charge density,
${\Delta^{E}_{(9-p)}} (r)$
is the Euclidean scalar propagator in $(9-p)$ transverse dimensions and
$V_{(p)}$ is the $p$-dimensional normalized volume. With $\rho_{(p)}^2
=T^{2}_{(p)}$ for
parallel branes, the above interaction energy vanishes.

For a D-brane - anti D-brane system the coupling to dilaton and graviton
is unchanged,
but the coupling to the RR tensor is reversed in sign, so that the
two terms in the interaction energy add.
Thus, the energy per unit volume is
\beq
\label{E}
E(r)=4\kappa^{2}_{(10)}T^{2}_{(p)}\Delta^{E}_{(9-p)}(r).
\eeq
The tension of a D-p brane is given by
$$T^{2}_{(p)}=\frac{\pi}{\kappa^{2}_{(10)}}\left(4\pi^2
\alpha^{'}\right)^{3-p},$$
where $\alpha^{'}=1/2 M_{s}^{-2}$ and  equation (\ref{E}) becomes
$$E(r)=4\pi (4\pi^2 \alpha^{'} )^{3-p} \Delta^{E}_{(9-p)} (r).$$

Let us now focus on D-3 branes.
Note that we could also consider alternative configurations, like a D-5
brane
wrapped on $S^2$ or any other effective D-3 branes.
This may alter the power law behavior of the interaction energy,
but would not dramatically influence the inflation scenario.
We have
$$E(r)=4\pi \Delta_{(6)}^{E}(r),$$
where
$$\Delta_{(6)}^{E}(r)=-\frac{1}{32\pi^3 r^4},$$
so that
\beq
E(r)=-\frac{1}{8\pi^2 r^4}.
\eeq
The total energy of a single pair of brane-antibrane is given by
\beq
V(r)=2 T_{(3)} + E(r)=2 T_{(3)}\left(1-\frac{1}{16\pi^2 T_{(3)}
r^4}\right).
\eeq

Let us now argue that such a setup provides us with a scalar (inflaton)
field.
The D-brane action is given by
\begin{eqnarray}
I_{D_{p}} &=& T_{(p)}\int d^{p+1}\zeta\;\; e^{(p-3)\Phi/4}
\sqrt{-\det{\hat{g}_{\alpha\beta }}} + ...
\nonumber\\
&\approx&
T_{(p)}\int d^{(p+1)}\zeta\;\; e^{(p-3)\Phi/4}\;
\frac{1}{2}\partial^{\alpha}X_{\mu}\partial_{\alpha}X^{\mu} + ... ,
\end{eqnarray}
where $X_{\mu}$ are the transverse coordinates. Because we have a parallel
brane - antibrane pair, we
can define $r=|X_{\mu}|$. So the D-brane action  gives us
a kinetic term for the scalar field, and in particular for D3-branes we
have
\beq
I_{(3)}=T_{(3)}\frac{1}{2}\int d^4 \zeta\;
\partial^{\alpha}X_{\mu}\partial_{\alpha}X^{\mu}\Rightarrow
\frac{1}{2}\int d^4 \zeta\; \partial^{\alpha}\phi\partial_{\alpha}\phi ,
\eeq
Such a scalar field corresponds to a small transverse oscillation of the
brane and, in fact, can be regarded as a Goldstone boson of the
spontaneously broken translation invariance in the extra space (by the
brane).
The inflaton field is a mode corresponding to a {\it relative} motion of two
such (parallel) branes and thus, is a
linear combination of two such scalars living on different
branes\cite{dva}.
The normalized  inflaton field $\phi =\sqrt{T_{(3)}} \; r$. In terms of
$\phi$
we have the following potential
\beq
\label{pot}
V(\phi)= 2 T_{(3)}\left( 1 - \frac{T_{(3)}}{16\pi^2 \phi^4}\right)\equiv
M^4\left(1-\frac{\alpha}{\phi^4}\right),
\eeq
with $\alpha =T_{(3)}/16\pi^2 $.

Following  $\cite{dva}$, we now explore the possibility of brane driven
inflation using the potential given in (\ref{pot}).  Let us first  outline
some of our assumptions:

(i) We assume that the extra dimensions are stabilized with
\beq
m^{2}_{Radion}\gg\ \frac {2T_{(3)}}{M^{2}_{P}}(\sim H^{2}).
\eeq
In such a situation the size of the extra dimensions can be
considered frozen during inflation.  The inter-brane motion
will drive four dimensional inflation.

(ii)  The effective four-dimensional Hubble size $H^{-1}$ should
be larger than the size $R_{c}$ of the extra dimensions.  This
allows us to treat the universe as four dimensional at distances
$\sim H^{-1}$.  The evolution of the four dimensional scale factor
is governed by
\beq
H^{2}=(\dot{a} / a)^{2}=\rho_{eff} / 3M^{2}_{P},
\eeq
where
$\rho_{eff}=T{\dot{r}}^{2}+V(r)$,
and we  require
$T{\dot{r}}^{2} \ll V(r)$.

For inflation we can consider more general potential
\beq
V(\phi)=M^4\left(1-\frac{\alpha}{\phi^n}\right),
\eeq
where $\phi$ satisfies the equation
\beq
\frac{d^2 \phi}{dt^2} +3 H \frac{d\phi}{dt} + \frac{dV}{d\phi}=0.
\eeq
To implement  inflation  we must satisfy the slow roll condition
$$\ddot{\phi} \ll 3 H\dot{\phi},$$
as well as the flatness conditions
\beq
\epsilon\ll 1,
\;\;\;\;\epsilon\equiv\frac{1}{2}M^{2}_{P}(\frac{V^{\prime}}{V})^{2},
\eeq
\beq
|\eta |\ll 1,\;\;\;\; \eta \equiv M^{2}_{P}\frac{V^{''}}{V}.
\eeq
The potential $V(\phi)$ from (\ref{pot})
is essentially flat with $\phi^n \gg \alpha$ which requires
\begin{eqnarray}
\phi &\gg& \left(\frac{1}{\sqrt{2}}M_{P}\alpha n\right)^{1/(n+1)},
\nonumber\\
\phi &\gg& \left(M_{P}^{2}\alpha n(n+1)\right)^{1/(n+2)}.
\end{eqnarray}
For $\alpha < M_{P}^{n}$, the second condition is
stronger, so the critical value of the field at which inflation
stops is given by
\beq
\phi_{c}=\left(M_{P}^{2}\alpha n(n+1)\right)^{1/(n+2)}.
\eeq

The number of $e$-foldings during inflation is given by
\beq
N(\phi)=\int_{\phi_{c}}^{\phi}  M_{P}^{-2} \frac{V(\phi)}{V^{'}(\phi)}
d\phi = \frac{1}{\alpha n M_{P}^{2}}\int_{\phi_{c}}^{\phi}
\left(\phi^{n+1} -\alpha\phi\right) d\phi\approx
\left(\frac{\phi}{\phi_{c}}\right)^{n+2}.
\eeq

For $n=4$ (D3-branes),
\beq
N(\phi)\approx\left(\frac{\phi}{\phi_{c}}\right)^{6}
=\frac{T^{3}_{(3)}r^6}{20\alpha M_{P}^2}=
\frac{4\pi^2 T^2_{(3)} r^6}{5 M_{P}^2}.
\eeq
The tension of the 3-brane is given by
$$T^2_{(3)}=\frac{\pi}{\kappa^2_{(10)}}=\pi M_{(10)}^8 = \pi
\frac{M_{P}^2}{V_{c}},$$
where $V_{c}$ is the volume of the compactified space.
Thus,
\beq
N(\phi)=\frac{4}{5}\pi^3 \frac{r^6}{V_{c}},
\eeq
where $r$ is the initial 3-brane - anti 3-brane separation.

Let us consider an example with a specific compactification geometry.
We will take $T^6$
with the same compactification radius $r_{c}$
for all dimensions, so that the maximal  initial distance between
the branes is $\sqrt{6}\pi r_{c}$.
For the number of $e$-foldings we have
\beq
N(\phi)=\frac{4}{5}\pi^3\frac{(\sqrt{6}\pi r_{c})^6}{(2\pi
r_{c})^6}\approx 80 ,
\eeq
so that inflation can be realized in this case.
It seems possible to increase the number of $e$-foldings
by choosing some other reasonable compactification geometry.

For the scalar spectral index one obtains
\beq
n\simeq 1-\frac{2}{N_{Q}}\simeq 0.97,
\eeq
which is in excellent agreement with the current measurements \cite{cmb}.
Furthermore, from the spectrum measurements by COBE  at the scale
$k\simeq 7.5H_{0}$ (more or less the center of the range explored
by COBE), and neglecting gravitational waves
(since $H\sim 10^{9}\mbox{GeV}\ll M_{P}$), one can deduce \cite{lyth}
$$ M_{p}^{-3}V^{3/2}/V^{'} =5.3\times 10^{-4}.$$
This gives us an estimate of the compactification volume $V_{c}$
\beq
M_{c}\equiv \left(V_{c}\right)^{-1/6}\sim 10^{12} \mbox{GeV}.
\eeq

In our discussion we have so far ignored the presence of the tachyonic
field $\chi$ associated with the brane-antibrane system \cite{sen}.
As long as they
are sufficiently far apart this is a reasonable approximation, and the
inflationary potential in (\ref{pot}) provides a very good approximation.
When
the brane-antibrane separation becomes comparable to the string scale,
however,
the field $\chi$ cannot be ignored. Let us parameterize the situation as
follows:
\beq
\label{cor}
V +\delta V = M^4 \left( 1- \frac{\alpha}{\phi^4}\right) + \chi^2 ( \phi^2
- M_s^2).
\eeq
Thus, as long as $\phi\gg M_s$ ( string scale), we can safely ignore the
second
term in (\ref{cor}), and the inflationary potential in (\ref{pot}) remains
valid. (Note that
$M_s\ge M_c$ ensures that $\phi_c \ge \phi_s$, where $\phi_s$ denotes the
value of $\phi$
at the string scale $M_s^{-1}$). However, for small values of $\phi$ the
tachyonic instability sets in and branes annihilate. This is the equivalent of
tachyon condensation into the vacuum. In this respect our brane inflation
is in some sense the hybrid inflation\cite{linde}\cite{shafi}\footnote{We thank E.
Kiritsis for this comment.}.

Finally, let us provide an order of magnitude estimate of the reheat
temperature
$T_r$ which appears after the merger and subsequent annihilation of the
brane-
antibrane system. We assume that the annihilation energy is dumped in the
bulk,
so that the effective decay rate $\Gamma$ on our 3-brane is
$\Gamma\sim T_{(3)}^{1/4}/(M_{10}R_{c})^6$, where $R_{c}\sim
10^{-12}\mbox{GeV}^{-1}$
denotes the effective length scale of the extra dimensions. The
oscillations are damped
out when the Hubble time becomes comparable to $\Gamma^{-1}$, and our
brane `reheats' to
a temperature $T_{r}\sim 0.1(\Gamma M_{P})^{1/2}\sim 10^{10}$GeV.

In conclusion, our brane inflationary scenario is based on a brane -
antibrane  system initially separated in the extra dimension and slowly moving
towards each other.
Due to this slow motion (protected by locality in the extra
space\cite{dva}) it can
circumvent the problem of 'slow roll' inflation encountered in SUGRA and
non-supersymmetric 4D-field theory models. The model can be naturally
applied to brane world models, and it also
works for bulk `inhabitants' with compactified extra dimensions. We find
an inflationary potential
of the form $V(\phi)=M^4 (1-\alpha/\phi^n),\;\;\; n>0$, which is a
hallmark of this class of
models. No fine tuning or small couplings are required. For inflation to
occur, the initial
brane - antibrane pair should be sufficiently far apart, but they need not
be exactly
(or almost exactly) parallel. All that is needed is that in a region of the
size of a Hubble volume the branes are approximately parallel so that the
gradient energy of the inter-brane separation field $\phi$ is subdominant
compared to the potential energy of separated branes. Such a region will then
expand exponentially and dominate the universe. In this respect, the
required initial conditions are identical to the ones in 4D hybrid
inflationary scenarios\cite{linde},\cite{shafi}.

 The magnitude of the scalar density fluctuations are proportional
to $M_{c}/M_{P}$, where $M_{c}\sim 10^{12}$GeV denotes the effective
compactification
scale of the six extra dimensions. The precise predictions are model
dependent and require a better
understanding of brane - antibrane annihilation dynamics, which should
also provide insights on
the gravitino constraint.
\\

{\bf Acknowledgments}
\\

We would like to thank E. Kiritsis and   M. Porrati for helpful
discussions.
The work of GD was supported in part by David and Lucille Packard
Foundation Fellowship for Science and Engineering, by Alfred P. Sloan
foundation fellowship
and by NSF grant PHY-0070787. The work of QS was supported by DOE under
contract DE-FG02-91ER40626.
\\

{\bf Note Added}
\\

The work described here was presented at a workshop in Heidelberg (April
4-7, 2001),
and also at the EURESCO conference in Le Londe (France) May 11-16, 2001.
After this talk was presented at the latter conference we
learned from F. Quevedo of similar work carried out by
C.P. Burgess, M. Majumdar, D. Nolte, F. Quevedo, G. Rajesh and R.-J.
Zhang.

\end{document}